# Near-field enhanced optical tweezers utilizing femtosecond-laser nanostructured substrates

**D.G. Kotsifaki,[1] M. Kandyla,[1,*] and P.G. Lagoudakis[2]**

[1]*Theoretical and Physical Chemistry Institute, National Hellenic Research Foundation, 48 Vasileos Constantinou Avenue, 11635 Athens, Greece*

[2]*Department of Physics and Astronomy, University of Southampton, Southampton, SO17 1BJ, UK*

**Abstract**

We present experimental evidence of plasmonic-enhanced optical tweezers, of polystyrene beads in deionized water in the vicinity of metal-coated nanostructures. The optical tweezers operate with a continuous wave near-infrared laser. We employ a Cu/Au bilayer that significantly improves dissipation of heat generated by the trapping laser beam and avoid de-trapping from heat convection currents. We investigate the improvement of the optical trapping force, the effective trapping quality factor, and observe an exponential distance dependence of the trapping force from the nanostructures, indicative of evanescent plasmonic enhancement.

Following the pioneering work of Ashkin *et al.*[1,2], the field of optical trapping and optical manipulation has experienced remarkable development, due to important applications such as cell sorting[3], investigating the DNA mechanics[4], observing the angular momentum of light[5], and probing the micro-rheological properties of particles[6], among others. While conventional optical trapping by a tightly focused laser beam has been extensively applied for the manipulation of micrometer-size particles[1,7,8] and biological samples[2,9], the trapping efficiency is limited for nanometer-size particles[10,11], mainly due to the diffraction limit of the trapping laser beam. To overcome this limitation, new techniques have been developed, which combine optical tweezers with novel trapping substrates, such as slot waveguides[12], ring resonators[13,14], photonic crystal resonators[15] and plasmonic nanostructures[10].

Plasmonic nanostructures yield deep sub-wavelength confinement of light and resonant enhancement of the optical field intensity[10], resulting in stable trapping[16-18] and enhanced optical forces[17,19,20]. Nanometric tweezers with sub-diffraction-limited trapping volume and increased efficiency have been developed by employing electromagnetically coupled pairs of gold nanopillars[20]. A periodic gold nano-antenna substrate, based on localized and extended surface plasmons, was proposed for optical trapping, nano-spectroscopy, and biosensing applications[21]. Localized surface plasmon-based (LSP) optical trapping of nanometric semiconducting quantum dots with weak light irradiation was achieved by gold nanodimer arrays[22]. Other implementations of LSP-based optical trapping include gold disks[23], gold stripes[24], arrays of gold nanoblock pairs[25], and arrays of gold micropads[26] as trapping substrates. Recently, optical trapping of polystyrene nanobeads by a two-dimensional lattice of gold nanostructures demonstrated that plasmonic optical lattices can be used to guide and arrange nanoparticles[27]. Apart from gold, silver is also an effective metal to obtain LSP resonant modes with high scattering efficiency[28]. However, because silver is prone to oxidation, there are only a few reports on silver-based plasmonic tweezers[29,30]. Although LSP tweezers are widely used for the manipulation of nanoscale objects, heating and thermal convection in the medium in which trapping is performed, associated with Ohmic losses in the metallic substrates, may affect the trapping process. This motivates the development of plasmonic nanotweezers employing thermal management. Hence, by using plasmonic nanopillars incorporating a heat sink, which transfers the generated heat away from the trapping volume, Wang *et al.* demonstrated stable trapping of 110-nm polystyrene beads[31].

Here, we employ a laser fabrication method for the development of plasmonic optical tweezers and investigate the enhancement of the optical trapping force (OTF) and the effective trapping quality factor. In addition, we systematically investigate the dependence of the trapping force on the distance from the substrate. The optical tweezers operate with a continuous wave (CW) near-infrared trapping laser beam and are based on femtosecond-laser nanostructured silicon samples, coated with thin bilayers of Cu/Au (see SI[32]). The OTF and effective trapping quality factor, obtained with Cu/Au-coated nanostructured silicon substrates, are one order of magnitude higher than those obtained with conventional optical tweezers in the absence of the nanostructure templates. The Cu/Au coating provides sufficient thermal transport away from the vicinity of the optical trap, allowing for measurements near the substrate surface. The OTF decays exponentially away from the Cu/Au-coated nanostructured silicon substrate, suggesting an enhancement mediated from the evanescent plasmon field. Uncoated nanostructured silicon samples were used as reference trapping substrates.

For optical trapping measurements, we employed a home-built trapping setup[33], with a near-infrared CW fiber laser operating at 1070 nm (Fig. 1). The trapping laser beam is focused by a high numerical aperture, oil-immersion microscope objective lens (NA = 1.4, x63) near the nanostructured silicon substrate. The focal point of the trapping laser beam has a diameter of $2\lambda/(\pi \, NA_{eff}) \approx 563$ nm (where $NA_{eff} = 1.21$)[34]. Fluorescent polystyrene beads of 400 nm diameter are trapped in deionized water near the focal point of the near-infrared laser beam. A weak probe CW laser beam of 474.3 nm wavelength and 1.32 mW power after the objective lens is used to induce fluorescence from the polystyrene nanobeads, which is imaged onto a CCD camera.

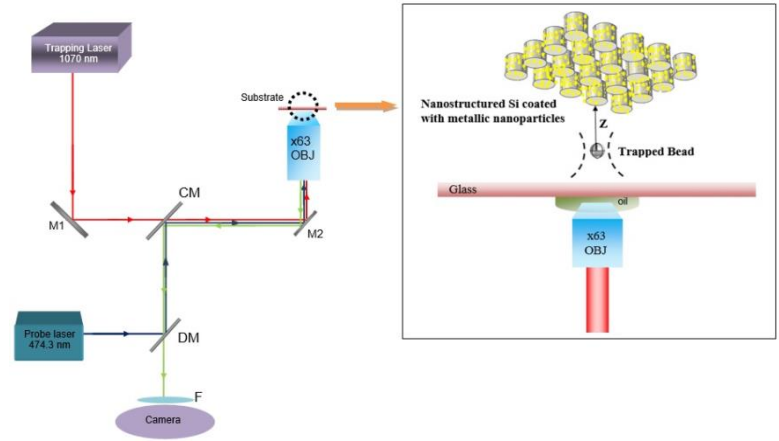

**Figure 1:** Schematic of the optical trapping setup. M1, M2 = mirrors; CM = cold mirror; DM = dichroic mirror; OBJ = microscope objective lens; and F = optical filters. The inset shows a close-up illustration of the optical trap above the nanostructured silicon substrate, coated with metallic nanoparticles. The bead is trapped at a distance $z$ above the substrate surface.

We were able to measure the distance of the optical trap from the surface of the trapping substrate by observing the speckle reflection of the trapping laser beam from the substrate and setting this position as the zero value of the distance with an accuracy of ±500 nm. The relative changes of the distance, $z$, with respect to the zero position are then determined with Rayleigh length resolution. We performed trapping measurements for various distances above different substrates and calculated the corresponding OTF for each distance. The OTF was determined by escape velocity measurements[35]. The trapping substrates were placed on a piezo-controlled stage, on which a continuous triangular function with a variable frequency was applied[33]. By measuring the minimum frequency for which a trapped bead escaped the optical trap, we were able to calculate the escape velocity and from that the OTF, $F$, using the modified Stokes law,

$$F = K6\pi\eta r v_{esc} \qquad (1)$$

where $\eta$ is the water viscosity, $r$ the bead radius, $v_{esc}$ the escape velocity of the bead, and $K$ is a dimensionless correction coefficient, determined by Faxen´s law[35]. Each escape velocity value resulted from the average of ten independent measurements. During optical trapping measurements, we systematically verified that the trapped beads were not mechanically pinned on the substrates, as they were released when switching off the trapping laser.

Figure 2 shows a scanning electron microscope image of a femtosecond-laser nanostructured silicon sample before metallic layer deposition. The structuring process results in a quasi-ordered distribution of columnar nanospikes on the silicon surface, with a mean tip diameter of ~175 nm and height ~360 nm. The average distance between neighbouring spikes is ~228 nm. Femtosecond-laser irradiation of silicon in liquid environments is known to generate nanospikes on the surface of the material due to ultrafast melting and interference effects[36,37].

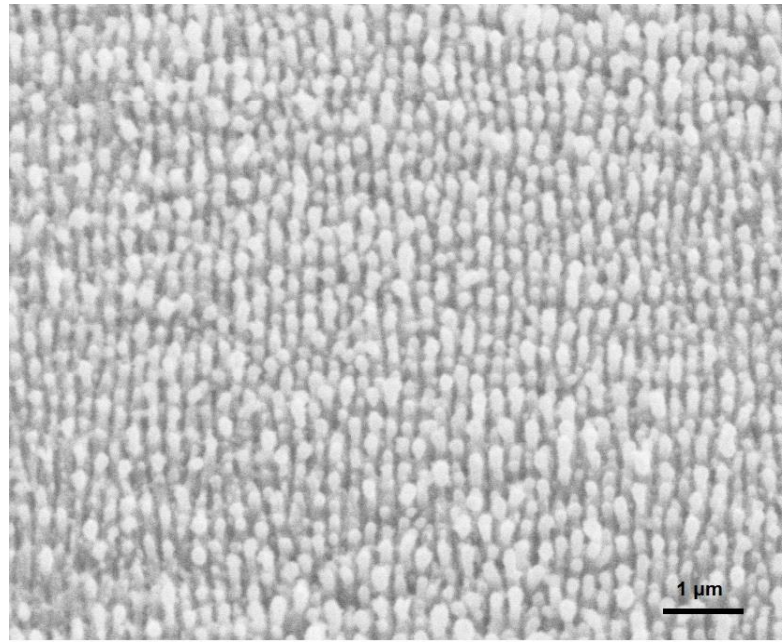

**Figure 2:** Scanning electron microscope image, viewed at 45° from the surface normal, of an uncoated femtosecond-laser nanostructured silicon sample.

Figure 3 shows the OTF exerted on 400-nm polystyrene beads, for various distances, $z$, above an uncoated flat (Fig. 3a) and an uncoated nanostructured (Fig. 3b) silicon substrate, as a function of the trapping laser power. The OTF above the flat silicon substrate does not show a systematic variation with the distance from the substrate. On the other hand, from Fig. 3b we notice the trapping force increases as the optical trap approaches the uncoated nanostructured silicon substrate. A plausible explanation for the OTF increase is light scattering or trapping from the silicon nanospikes. It has been reported that silicon nanowire arrays effectively trap light due to optical scattering in the plane of the sample[38-40]. This effect, which is beyond the scope of the current paper, is the subject of future studies.

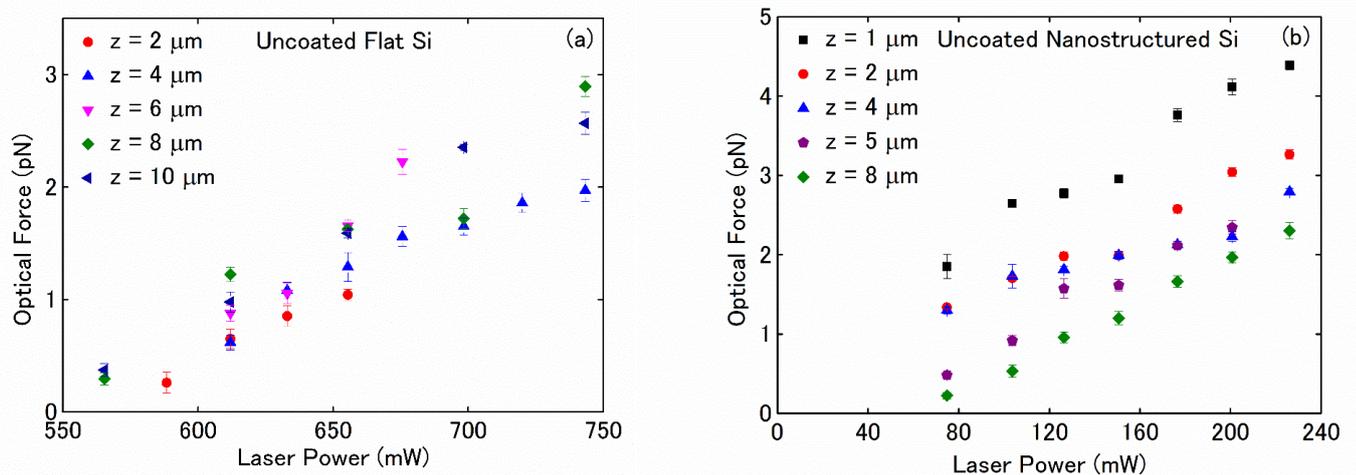

**Figure 3:** Optical trapping force as a function of the trapping laser power for (a) an uncoated flat silicon and (b) an uncoated nanostructured silicon substrate, for various distances, $z$, above the substrates. The y-error corresponds to the standard deviation of the optical force measurement.

In order to compare the performance of the two substrates, we calculate the effective trapping quality factor, $Q$, for each substrate, according to the definition

$$Q = Fc/nP \qquad (2)$$

where $F$ is the OTF, $c$ is the speed of light, $P$ is the trapping laser power, and $n$ is the refractive index of the medium containing the trapped beads ($n = 1.326$ for deionized water). For the flat silicon substrate at $z = 2$ μm above the surface, the effective quality factor varies from $Q = (9.9 \pm 3.5) \times 10^{-5}$ to $Q = (3.63 \pm 0.17) \times 10^{-4}$ with an average value of $Q_{avg} = (2.5 \pm 1.1) \times 10^{-4}$. For the nanostructured silicon substrate at $z = 2$ μm above the surface, the effective quality factor varies from $Q = (2.990 \pm 0.071) \times 10^{-3}$ to $Q = (4.040 \pm 0.067) \times 10^{-3}$ with an average value of $Q_{avg} = (3.47 \pm 0.34) \times 10^{-3}$. We observe the effective quality factor for the uncoated nanostructured silicon substrate is one order of magnitude higher than the effective quality factor for the flat silicon substrate.

Figure 4 shows the OTF exerted on 400-nm polystyrene beads, for various distances, $z$, above a flat silicon substrate coated with a Cu/Au layer (Fig. 4a) and a nanostructured silicon substrate coated with a Cu/Au layer (Fig. 4b), as a function of the trapping laser power. Comparing Figures 4a and 4b, we notice the OTF increases in the presence of the silicon nanostructure for the Cu/Au-coated substrates. The effective trapping quality factor we obtain for the Cu/Au-coated flat silicon substrate at $z = 1$ μm above the surface varies from $Q = (2.56 \pm 0.11) \times 10^{-3}$ to $Q = (12.75 \pm 0.75) \times 10^{-3}$ with an average value of $Q_{avg.} = (9.2 \pm 3.8) \times 10^{-3}$. The increase of the quality factor of the Cu/Au-coated flat silicon substrate with respect to the uncoated flat silicon substrate could be due to the excitation of a plasmon mode, supported by the roughness or discontinuity of the thermally evaporated thin Cu/Au film[41-44]. For the Cu/Au-coated nanostructured silicon substrate at $z = 1$ μm above the surface, the quality factor varies from $Q = (9.05 \pm 0.29) \times 10^{-2}$ to $Q = 0.1340 \pm 0.0039$ with an average value of $Q_{avg} = 0.117 \pm 0.016$. The quality factor for the Cu/Au-coated nanostructured silicon substrate is one order of magnitude higher than the quality factor for the Cu/Au-coated flat silicon substrate and 24 times higher than the quality factor for the uncoated nanostructured silicon substrate ($Q_{avg}$ = (4.95 ± 0.55)x10$^{-3}$ at $z$ = 1 $\mu$m).

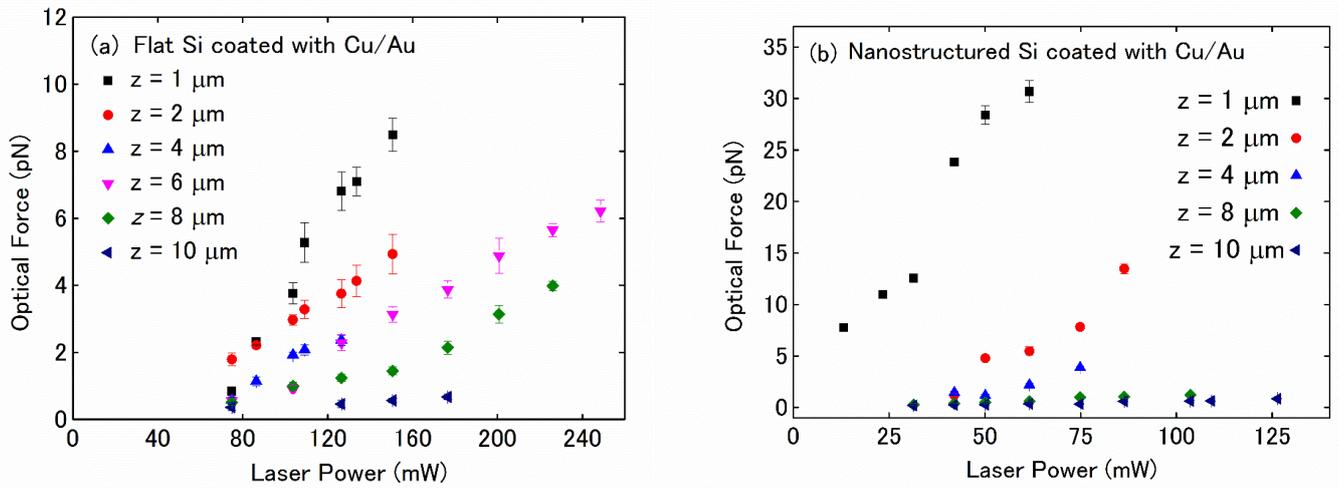

**Figure 4:** Optical trapping force as a function of the trapping laser power for (a) a flat silicon substrate coated with 3-nm copper/50-nm gold and (b) a nanostructured silicon substrate coated with 3-nm copper/50-nm gold, for various distances, $z$, above the substrates. The y-error corresponds to the standard deviation of the optical force measurement.

We also obtained OTF measurements with silver-coated flat and nanostructured silicon substrates (see SI[32]). However, we observe that for silver-coated nanostructured silicon substrates, bubble formation does not allow measurements for distances less than z = 3 $\mu$m from the substrate surface. The formation of bubbles is associated with thermal absorption[45] or dielectric breakdown in a liquid[46] or electromagnetic stress[47]. Thermal absorption requires light intensities of ~10$^5$ W/cm$^2$ for bubble formation[45]. Here, the intensity of the focused trapping beam was ~10$^6$ W/cm$^2$, which exceeds the intensity required for thermal absorption. However, we did not observe bubble formation for the Ag-coated flat substrate. This implies that thermal absorption is not the only mechanism responsible for bubble formation. In order to avoid bubble formation and perform trapping measurements closer to coated nanostructured silicon substrates, we employed a Cu/Au coating, which results in suppression of liquid instabilities and convective effects[31].

Figure 5 summarizes the distance dependence of the trapping force for the Cu/Au-coated and uncoated nanostructured substrates. It shows the OTF exerted on 400-nm polystyrene beads, obtained with an uncoated nanostructured silicon substrate for a trapping laser power of 74.9 mW and the trapping force obtained with a Cu/Au-coated nanostructured silicon substrate, for trapping laser powers of 50.1 mW and 61.6 mW, as a function of the relative distance with respect to the zero position, $z$, between the trapping laser beam focus and the substrate. Due to unstable optical trapping, there are no data available for a trapping laser power of 50.1 mW or 61.6 mW for the uncoated nanostructured silicon substrate, therefore we present the nearest trapping laser power value for which we were able to acquire measurements. The OTF obtained with the uncoated nanostructured silicon substrate varies only slightly with the distance from the substrate surface. On the contrary, the OTF varies strongly with the distance from the substrate for the nanostructured silicon substrate coated with Cu/Au, especially for distances smaller than $z = 4$ $\mu$m.

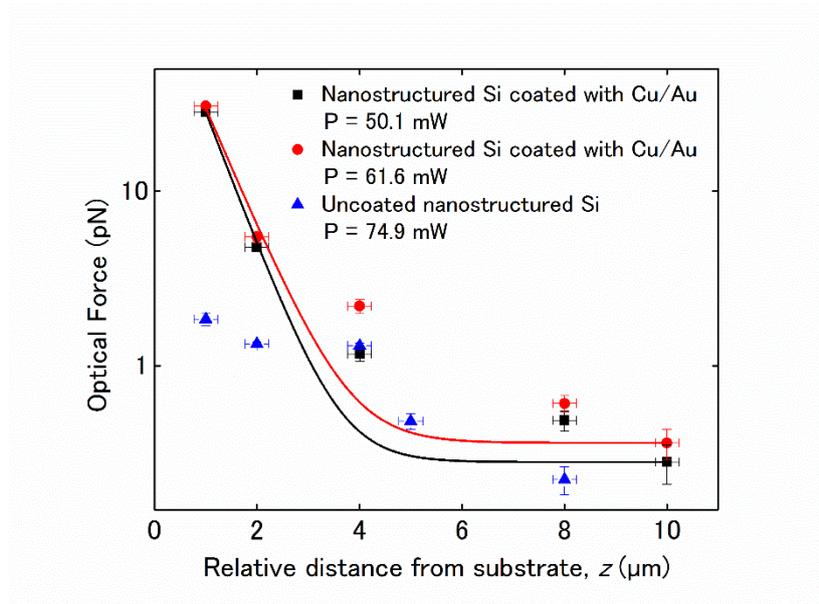

**Figure 5:** Semilog plot of the optical trapping force as a function of the relative distance, *z*, above the trapping substrate, for coated and uncoated nanostructured silicon substrates. Relative distance is the measured distance with respect to the zero position. The trapping laser power was 74.9 mW for the uncoated nanostructured silicon substrate, 50.1 mW and 61.6 mW for the Cu/Au-coated nanostructured silicon substrate. Solid lines: exponential fit to the data obtained with the Cu/Au-coated nanostructured silicon substrate. The x-error corresponds to the Rayleigh length of the trapping laser beam and the y-error to the standard deviation of the optical force measurement.

The solid lines in Figure 5 represent a fit to the data, according to the equation:

$$F = F_o + A e^{-bz} \qquad (3),$$

where $F_o$ is the OTF value at the distance $z = 10$ $\mu$m above the coated nanostructured silicon substrate for each laser power (50.1 mW and 61.6 mW) and *A, b* are fitting parameters. The trapping force obtained with the Cu/Au-coated nanostructured silicon substrate decays exponentially with the distance from the substrate, as shown by the solid lines in Fig. 5, with a decay length of 600 ± 30 nm. The decay length of an evanescent plasmon field is typically on the order of half the wavelength involved[47]. In this work, the wavelength of the trapping laser beam is 1070 nm.

The enhancement of the OTF and effective quality factor, induced by the Cu/Au-coated nanostructured silicon substrate, combined with the exponential decay of the trapping force away from the substrate, are attributed to the excitation of localized surface plasmon modes, which results in an enhancement of the electromagnetic field in the vicinity of the optical trap. Deposition of thin metallic layers on silicon surfaces with nanometric roughness is known to favor the formation of metallic nanoparticles, instead of a smooth metallic film[48]. Microstructured silicon samples with a thin silver coating, employed as optical trapping substrates, have also shown two orders of magnitude enhancement of the quality factor ($Q = 0.0540 \pm 0.0016$ at $z = 1$ $\mu$m) with respect to uncoated trapping substrates[30]. For the Cu/Au-coated nanostructures studied here, we observe an even higher quality factor of ~ 0.13, similar to the highest reported quality factor for plasmon-enhanced optical traps[20].

Even though several studies have observed the enhancement of the OTF above plasmonic substrates of different geometries[10], the dependence of the OTF on the distance from the substrate is rarely reported. Grigorenko *et al.* measured the OTF above pairs of gold nanopillars as a function of the distance from the nanopillars and report qualitatively that the trapping force increases dramatically for small distances[20]. Also, Volpe *et al.* measured the radiation force, exerted on trapped dielectric particles, as a function of the distance from a thin gold layer and report that the radiation force decays away from the substrate[17]. Here, we systematically measure the OTF above several substrates as a function of the distance from the substrate, for various trapping laser powers, and observe that for the Cu/Au-coated nanostructured silicon substrate the trapping force decays exponentially away from the substrate. The exponential decay of the trapping force indicates the presence of an evanescent electromagnetic field, which may stem from the excitation of localized surface plasmon modes or from evanescent coupling with waveguided modes. This study, combined with Refs.[17,20], puts strong evidence of a plasmonic enhancement. However, conclusive evidence would require a wavelength dependence of the coupling strength, which is the subject of a future study.

It is important to stress that the trapping data presented in this work have been obtained in the absence of local heating effects. We always performed measurements at laser powers and distances from the substrates where thermal convection was not observed. Furthermore, we note that the dependence of the OTF on the trapping laser power is always linear, indicating the absence of significant heating in the surrounding medium. We observed the Cu/Au coating results in the absence of bubble formation and convection, as opposed to the silver coating. The combination of trapping force enhancement with the absence of liquid instabilities and convective effects, allows for working in the proximity of this trapping substrate and taking full advantage of its near-field properties for optical trapping experiments and applications.

In conclusion, we demonstrate an optical tweezer system, based on femtosecond-laser nanostructured silicon substrates, coated with thin metallic layers, which shows an order of magnitude enhancement of the OTF and the effective quality factor, compared with unstructured substrates. We try two different metallic coatings, silver and Cu/Au. The Cu/Au coating results in suppressed liquid instabilities and convective effects, allowing for trapping closer to the substrate surface. We observe an exponential decay of the trapping force away from the Cu/Au-coated nanostructured substrate, indicative of a plasmon-enhanced optical tweezer. Trapping force enhancement and efficient thermal management results in efficient trapping of nanoparticles, allowing for the manipulation of biological samples with optical tweezers while avoiding photodamage.

**Acknowledgements**

Financial support of this work by the General Secretariat for Research and Technology, Greece, (project Polynano-Kripis 447963) is gratefully acknowledged.

# Near-field enhanced optical tweezers utilizing femtosecond-laser nanostructured substrates

**Supplemental Material**

**Fabrication process**

We prepared the nanostructured substrates by irradiating silicon samples with trains of femtosecond laser pulses at normal incidence in distilled water[S1]. An amplified Ti:sapphire laser system was used to generate 800-nm center wavelength, 200-fs pulses at a repetition rate of 20 KHz. The pulses were frequency-doubled to a center wavelength of 400 nm using a $BBO_3$ crystal. The average power of the pulse train after the $BBO_3$ crystal was 25 mW. The laser pulses were focused by a 10x objective lens on the silicon wafer surface, to a fluence of ~1 J/cm². The silicon wafer was placed in a cuvette filled with distilled water and irradiated by approximately 60.000 pulses. The nanostructured silicon samples were coated either with a 50-nm silver layer or with a 3-nm copper layer followed by a 50-nm gold layer (Cu/Au) by thermal evaporation, in order to be employed as plasmonic substrates for optical trapping.

**Optical trapping force exerted on beads above Ag-coated silicon substrates**

Figure S1 shows the optical trapping force exerted on 400-nm polystyrene beads, for various distances, $z$, above a flat silicon substrate coated with a 50-nm silver layer (Fig. S1a) and a nanostructured silicon substrate coated with a 50-nm silver layer (Fig. S1b), as a function of the trapping laser power. In both cases, the optical trapping force increases closer to the substrate. For the Ag-coated nanostructured silicon substrate the optical trapping force increases abruptly for $z = 3$ $\mu$m (Fig. S1b). The inset in Fig. S1b shows the trapping force for distances $z = 4, 6, 8, 10$ $\mu$m, plotted on a finer scale. For this substrate, we were not able to acquire trapping data for distances less than $z = 3$ $\mu$m, due to bubble formation, even for low trapping laser powers. The effect of thermal convection was not as intense for the Ag-coated flat silicon substrate, for which we were able to acquire trapping data for distances as low as $z = 1$ μm (Fig. S1a). The effective quality factor, obtained with the Ag-coated flat silicon substrate at $z = 1$ μm above the surface, varies from $Q = (1.047 \pm 0.048) \times 10^{-2}$ to $Q = (1.544 \pm 0.086) \times 10^{-2}$ with an average value of $Q_{avg} = (1.20 \pm 0.18) \times 10^{-2}$. The effective quality factor, obtained with the Ag-coated nanostructured silicon substrate at $z = 3$ μm above the surface, varies from $Q = (5.236 \pm 0.036) \times 10^{-2}$ to $Q = (9.825 \pm 0.022) \times 10^{-2}$ with an average value of $Q_{avg} = 0.071 \pm 0.020$. Therefore, the presence of the silicon nanostructure improves the quality factor by a factor of six for the Ag-coated substrates. We note the above comparison is performed for different distances above the trapping substrates, indicating the enhancement factor would probably be higher if we were able to acquire trapping data closer to the Ag-coated nanostructured substrate.

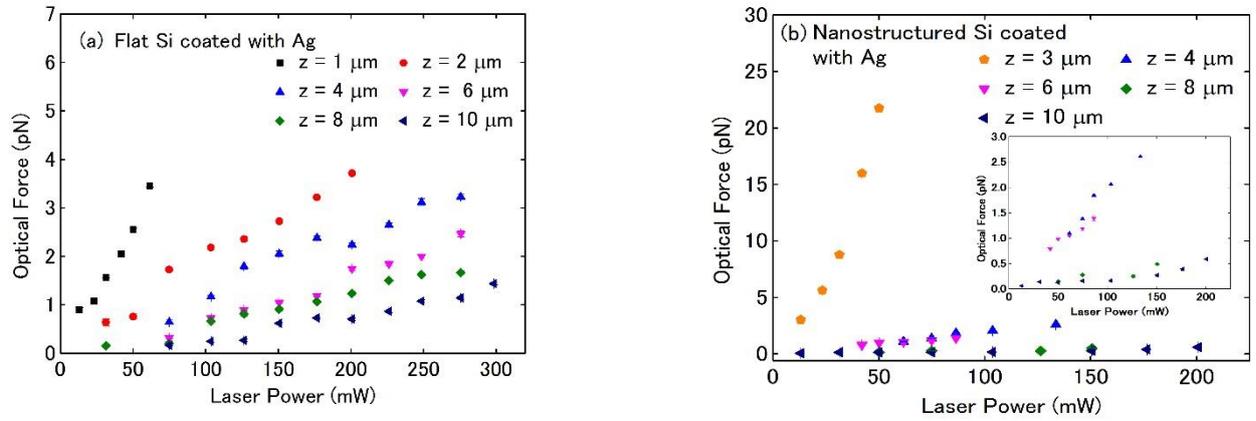

Figure S1: Optical trapping force as a function of the trapping laser power for (a) a flat silicon substrate coated with a 50-nm silver layer and (b) a nanostructured silicon substrate coated with a 50-nm silver layer, for various distances, $z$, above the substrates. The y-error corresponds to the standard deviation of the optical force measurement.